\documentclass[referee]{aa}
\usepackage{epsfig}
\usepackage{graphicx}

\usepackage{epstopdf}
\newcommand{\eqref}[1]{(\ref{#1})}

\begin{document}

\title{Transverse kink oscillations of expanding coronal loops}
\author{ Istvan Ballai  \and Beniamin Orza } 
\institute{Solar Physics and Space Plasma Research Centre (SP$^2$RC), Department of
Applied Mathematics, The University of Shef{}field, Shef{}field,
UK, S3 7RH, email: {\tt\{b.orza;i.ballai\}@sheffield.ac.uk}}
\date{Received dd mmm yyyy / Accepted dd mmm yyyy}

\authorrunning{Ballai and Orza}
\titlerunning{Transverse kink oscillations of expanding coronal loops}

\abstract{}
{We investigate the nature of transverse kink oscillations of loops expanding through the solar corona and how can oscillations be used to diagnose the plasma parameters and the magnetic field. In particular, we aim to analyse how the temporal dependence of the loop length (here modelling the expansion) will affect the $P_1/P_2$ period ratio of transverse loop oscillations.}
{Due to the uncertainty of the loop's shape through its expansion, we discuss separately the case of the loop that maintains its initial semi-circular shape and the case of the loop that from a semi-circular shape evolve into an elliptical shape loop. The equations that describe the oscillations in expanding flux tube are complicated due to the spatial and temporal dependence of coefficients. Using the WKB approximation we find approximative values for periods and their evolution, as well as the period ratio.  For small values of time (near the start of the expansion) we can employ a regular perturbation method to find approximative relations for eigenfunctions and eigenfrequencies. }
{Using simple analytical and numerical methods we show that the period of oscillations are affected by the rising of the coronal loop. The change in the period due to the increase in the loop's length is more pronounced for those loops that expand into a more structured (or cooler corona). The deviation of periods will have significant implications in determining the degree of stratification in the solar corona. { The effect of expansion on the periods of oscillations is considerable only in the process of expansion of the loop but not when it reached its final stage}.}
{The present study improves our understanding of the complexity of dynamical processes in the solar corona, in particular the changes of periods of kink oscillations due to temporal changes in the characteristics of the coronal loop. Our results clearly show that the problem of expansion of coronal loops can introduce significant changes in the period of oscillations, with consequences on the seismological diagnostics of the plasma and magnetic field. }

\keywords{Magnetohydrodynamics (MHD)---Sun: corona---Sun: magnetic fields---Sun: oscillations}

\maketitle

\section{Introduction}

Dynamical processes and transients observed in solar and space plasmas received considerable attention due to their ability to help scientists to diagnose remotely the parameters of the medium (temperature, scale-height, density, transport coefficients, etc.) in which these events occur, the magnitude and structure of the magnetic field, and the stability of the plasma. Seismological techniques and methods, imported from Earth's seismology and helioseismology, assume the combination of high resolution observations (amplitude, wavelength, propagation speed, damping time/length) with theoretical models (dispersion and evolutionary equation) in order to derive quantities that cannot be measured directly or indirectly. In particular, coronal seismology emerged as one of the most dynamically developing methods of solar physics  (see, e.g. 
Roberts et al. 1984, Nakariakov et al, 1999. Ruderman and Roberts 2002, Andries et al. 2005, 2009, Ballai et al. 2005, Banerjee et al. 2007, Verth et al. 2007, Ballai 2007, Ruderman et al. 2008, Morton and Erd\'elyi 2009, Ruderman and Erd\'elyi 2009, Wang et al. 2012, just to name a few).

One important candidate in coronal seismology are transverse kink oscillations, i.e. oscillations which exhibit 
periodic movement about the loop's symmetry axis. Recent CoMP observations (Tomczyk et al. 2007) showed that the predominant motion of coronal loops is the transverse kink oscillation and this is the easiest to generate. The triggerring of these oscillations can be through a lateral forcing process at arbitrary height of the loop by a blast wave (or EIT wave) that emanates as a result of a sudden energy release by a flare and/or CME (see, e.g. Ballai 2007, Ballai et al. 2008). Secondly, kink oscillations can also be triggered by the transverse motion of the footpoints due to the granular buffeting of flux tubes in the photosphere. This scenario is true not only for coronal structures, but applies to all magnetic entities in the solar atmosphere that can serve as waveguides. 
Global EIT waves can also interact with prominence fibrils, as observed by, e.g. Ramsey and Smith (1966), and more recently by Eto et al. (2002), Jing et al. (2003), Okamoto et al. (2004), Isobe and Tripathi (2007), Pint\'er et al. (2008) (for a comprehensive review see, e.g. Arregui et al. 2012) in order to generate kink waves and oscillations in prominences.

The dispersion
relations for many simple (and some quite complicated) plasma
waves under the assumptions of ideal magnetohydrodynamics (MHD)
are well known; they were derived long before accurate EUV
observations were available (see, e.g. Edwin and Roberts 1983, Roberts et al. 1984) 
using simplified models within the framework of ideal and linear
MHD. Although the realistic interpretation of many observations are
made difficult by the spatial and temporal resolution of present
satellites not being quite sufficient, considerable amount of information
about the state of the plasma, and the structure and magnitude of the
coronal magnetic field, can still be obtained.

The mathematical description of waves and oscillations in solar structures is, in general, given by equations whose coefficients vary in space and time. It has been recognised by, e.g.  Andries et al. (2005) that the longitudinal stratification (i.e. along the longitudinal symmetry axis of the tube that coincides with the direction of the magnetic field) is modifying the periods of oscillations of coronal loops. Accordingly, in the case of kink waves, these authors showed that the ratio $P_1/P_2$ (where $P_1$ refers to the period of the fundamental transverse oscillation, while $P_2$ describes the period of the first overtone of the same oscillation) can differ - sometimes considerably - from the canonical value of 2, that would be recovered if the loops were homogeneous. These authors also showed that the deviation of $P_1/P_2$ from 2 is proportional to the degree of stratification. This problem was also discussed in other studies such as Dymova and Ruderman 2006, Diaz et al. 2007, McEwan et al. 2008, Ballai et al. 2011, Orza et al. 2012. In a recent analysis, Ballai et al. (2011) discusses about the ambiguity of the period ratio seismology, as some other effects could result in the observation of multiple periods and each interpretation results in different value for the magnetic field and/or degree of stratification.

Later, studies by, e.g. Verth et al. (2007), showed that it is not only density stratification that is able to modify the  $P_1/P_2$ period ratio, but the variation of the loop's cross section area has also an effect on the period ratio. While the density stratification tends to decrease the period ratio, a modification of the cross section (i.e. when the magnetic field is flaring up as we approach the apex) tends to increase the $P_1/P_2$ value.

The investigation of properties of oscillations of coronal loops when equilibrium parameters of the plasma depend on time is a relatively new area of dynamical studies in the solar atmosphere, in particular for coronal seismology. Morton and Erd\'elyi (209) and Morton et al. (2010) studied the effect of cooling (i.e. temporal dependence of temperature) on the dynamics of kink oscillations and travelling waves and they found that the cooling of the plasma results period decrease and amplification of oscillations and a mean to dissipate the energy stored in waves propagating in an unbounded plasma. The same idea was used later by Morton et al. (2011) to study the properties of torsional Alfv\'en waves in coronal loops. The problem was recently re-considered by Ruderman (2011a) who showed that the cooling also generates an amplification of kink oscillations. The amplification of oscillations appears to be a competing effect with the damping due resonant absorption as shown by Ruderman (2011b) 

The problem of loop emergence and expansion through the solar atmosphere is one of the most challenging topics of solar physics as it involves the analysis of the evolution of the magnetic field in different regions of the solar interior and atmosphere where conditions can change from region to region. According to the standard theory, the magnetic field produced  by the dynamo action in the tachocline is transported through the solar convective zone towards the solar surface by magnetic buoyancy coupled with convective motion (Parker 1955, 1988). Once at the surface,  the emerged flux tube creates sunspots and bipolar active regions (Zwaan 1987). In the solar atmosphere the rise of the flux tube continues due to an excess of the magnetic pressure inside the loop (e.g. Archontis et al. 2004). For the purpose of our investigation, we will assume that this excess is balanced at the transition region (TR) and from this height the expansion is not a driven problem any longer, instead the loop moves through the corona in the virtue of its inertia. During the emergence and expansion phase, the flux tube can interact with existing magnetic structure in the solar atmosphere and this might be responsible for the appearance of small-scale (e.g. compact flares, plasmoids, X-point brightenings) and large-scale events (flares and CMEs) as suggested by Archontis (2004). Loop emergence often is associated with strong upflows as observed by, e.g. Harra et al. (2010, 2012). 

It is rather straightforward to imagine what happens with oscillations in a loop when the length of the loop is increasing. As the length of the loop becomes larger, the frequency of oscillations becomes smaller, i.e. the periods of oscillations are expected to grow, however in an inhomogeneous waveguide particular periods will be differently affected by the combined effect of inhomogeneity and loop length increase. Therefore, we expect that the period ratio of oscillations will change not only with the degree of inhomogeneity but also with time.

The aim of this paper is to investigate the effect of the loop expansion through the solar corona on the period ratio $P_1/P_2$ and the consequences of the inclusion of the length of the loop as a dynamical parameter on estimations of the degree of density stratification. The paper is structured in the following way: in Section 2 we introduce the mathematical formalism and obtain analytical results for an expanding loop that initially starts from a semi-circular shape and this shape is preserved throughout the expansion.  Later, in Section 3 we generalise our findings by assuming that the expansion of the loop in the solar corona does not occur with the same speed in the horizontal and vertical direction, in this case, the initial semi-circular loop transforms into a loop with elliptic shape. Finally, our results are summarised in the last section.

\section{The mathematical formulation of the problem}

In our analysis we will capture the dynamical behaviour of the loop once reached the transition region (TR). At this height, the loop can interact with a blast wave propagating in the low solar corona. This instigator could be easily identified with an EIT wave that propagates in the low corona over very large distances and they are known to be one of the major source for kink oscillations of coronal loops (see, e.g. Ballai 2007). We assume that the height at which the loop starts its journey through the corona is at 3 Mm above the solar surface. A typical loop length is of the order of 300 Mm. For practical reasons, we are going to consider that the height of the loop in its final position would be about 97 Mm resulting in a loop length of about 305 Mm.

The raising speed of loops is generally taken to be 10-15 km s$^{-1}$ (Chou and Zirin 1988, Archiontis 2008), therefore the time needed for the loop to travel the distance from the TR to its steady position is easily estimated to be between 3.4 and 5.2 hours. This time is at least two order of magnitude larger than a typical period of kink oscillations, so there is enough time for the development of oscillations. Higher rising speeds are also possible, recently Schmidt and Ofman (2011) reported expansions of a post-flare loop with speeds of hundreds km s$^{-1}$. Standing waves are formed if the speed of change in the length of the loop is smaller than the period of oscillations. This condition is easily satisfied for fast kink oscillations. We assume that the loop expands into the "empty" corona, i.e. it will not encounter any interaction with existing magnetic elements.

In the first instance we assume that the semi-circular shape of the loop at the TR is preserved throughout the expansion. Assuming an initial height of 3 Mm above the surface, the distance between the footpoints of the loop is 6 Mm. When reaching the final height of 190 Mm, the footpoints travel over a distance of 197 Mm. In addition we assume that  the expansion occurs at constant temperature (isothermal process) and throughout the expansion the loop maintains its cross-section constant. Due to the increase in the volume of the loop, a pressure difference is generated meaning that the plasma flows along magnetic field lines resulting in a density that depends not only on the height, $z$, but also on time. As it raises through the solar corona, the tangent to the loop is also changing monotonically, therefore the momentum equation in an equilibrium state becomes
\begin{equation}
\rho_0\frac{\partial u}{\partial t}+\rho_0u\frac{\partial u}{\partial z}=-\frac{\partial p_0}{\partial z}-g\rho_0\cos\beta ,
\label{eq:1.1}
\end{equation}
where the angle $\beta$ is a function depending on time and space, $u$ is the equilibrium flow of the plasma, $\rho_0$ and $p_0$ are the equilibrium density and pressure, and $g$ is the gravitational acceleration. In coronal loops, the flows are of the order of a few tens of km s$^{-1}$, therefore in Eq. (\ref{eq:1.1}) we can neglect the terms on the left hand side since
\[
u\frac{\partial u}{\partial z}\sim \frac{u^2}{L}\approx \frac{10^8}{10^8}={\cal O}(1),
\]
that is two order of magnitude smaller than terms on the right hand side. We can write pressure as
\[
p_0(z,t)=\frac{k_BT_0\rho_0(z,t)}{m}
\]
where $k_B$ is the Boltzman constant, $T_0$ is the constant temperature and $m$ is the mean atomic mass per particle. Introducing this expression into the RHS of Eq. (\ref{eq:1.1}) we obtain that
\begin{equation}
\frac{1}{\rho_0}\frac{\partial \rho_0}{\partial z}+\frac{\cos \beta(z,t)}{H}=0,
\label{eq:1.2}
\end{equation}
where $H$ is the constant density scale-height. Focussing on the density distribution inside the loop, we can integrate the above equation to obtain
\begin{equation}
\rho_i=\rho_f\exp\left[-\int_0^z\frac{ \cos \beta(z',t)}{H}dz'\right].
\label{eq:1.3}
\end{equation}
Assuming that the loop is semicircular we obtain that $\beta(z,t)=\pi z/L(t)$, therefore the density inside the coronal loop becomes
\begin{equation}
\rho_i=\rho_f\exp\left[-\frac{L(t)}{\pi H}\sin\frac{\pi z}{L(t)}\right],
\label{eq:1.4}
\end{equation}
where $\rho_f$ is the density of the plasma at the footpoint of the loop. For simplicity we will assume that the external density can simply be written as $\rho_e=D\rho_i$, where throughout our calculations we consider $D=0.5$. Figure 1 depicts schematically the change of the equilibrium density, both in space and time.
\begin{figure}
\centering
\includegraphics[width=\columnwidth]{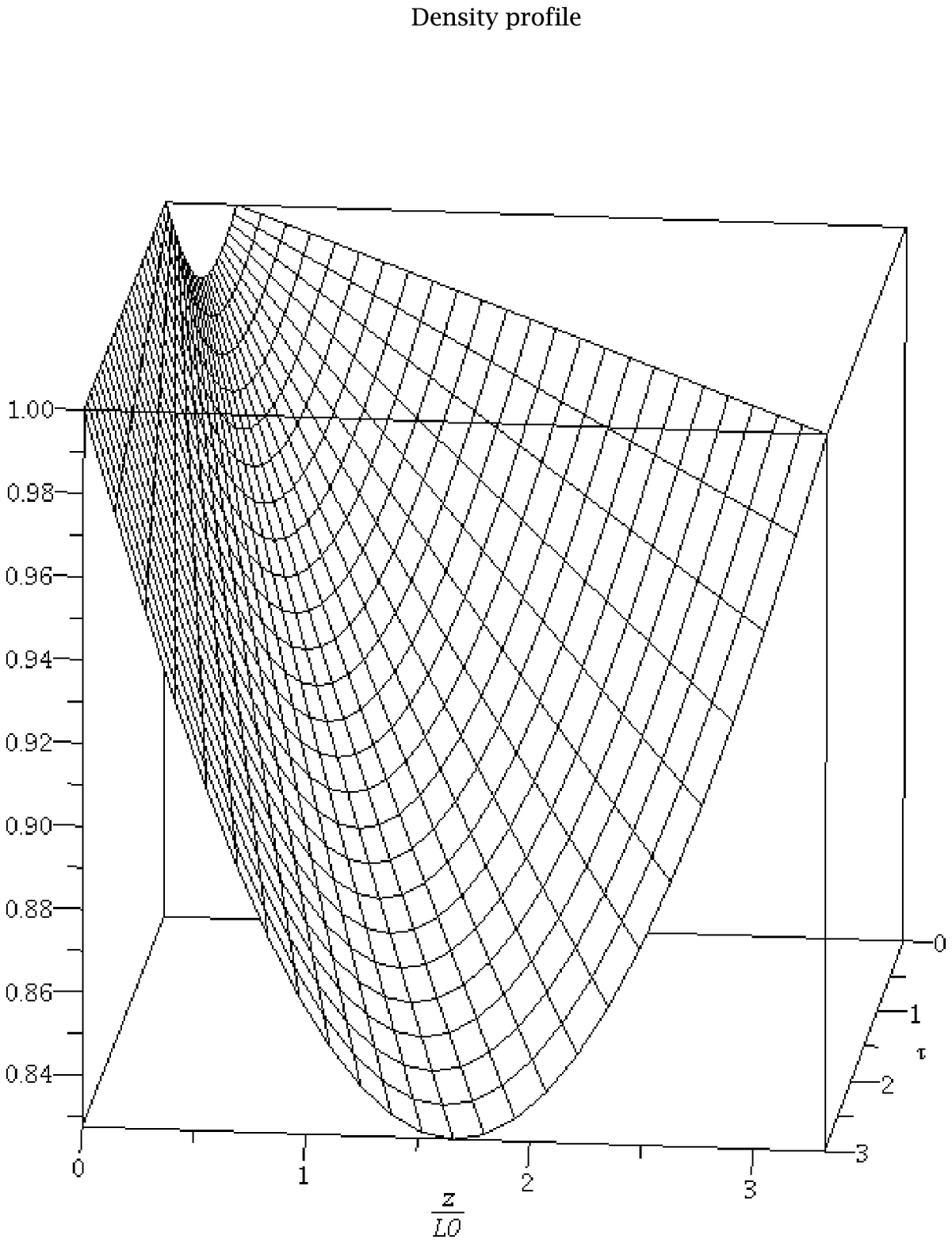}
\caption{A schematic representation of the evolution of the equilibrium density measured on the vertical axis in the units of density at the footpoint. Here lengths are given in units of the loop length at the start of the expansion in the corona ($L_0$) and time is given in units of $L_0/v_r$, where $v_r$ is the constant rising speed in the vertical direction, here taken to be 15 km $s^{-1}$}
\label{fig1}
\end{figure}
Here length was normalised to the length of the loop at the start of the emergence (considered $L_0=3\pi$ Mm) and time (here denoted by $\tau$) was normalised to the quantity $L_0/v_r$, where $v_r$ is the vertical rising speed of the loop considered $v_r=15$ km s$^{-1}$. According to our expectations, the density of the loop decreases with time. We assume that the flux tube is thin even at the beginning of its expansion.

In the thin flux tube approximation the dynamics of transverse kink oscillations is given by (see, e.g. Ruderman 2010, 2011a)
\begin{equation}
\rho_i\left(\frac{\partial }{\partial t}+U_i\frac{\partial }{\partial z}\right)^2\eta+ \rho_e\left(\frac{\partial }{\partial t}+U_e\frac{\partial }{\partial z}\right)^2\eta-\frac{2B^2}{\mu_0}\frac{\partial^2\eta}{\partial z^2}=0,
\label{eq:1.5}
\end{equation}
where $U_i$ and $U_e$ are the internal and external flows, $B$ is the magnetic field strength (here assumed to have identical value inside and outside the loop), $\mu_0$ is the magnetic permeability of the free space and $\eta$ is a complex valued displacement of the loop, with $\eta=\eta_R+i\eta_I$. In Cartesian coordinates the loop's displacement in the $x$ and $y$ directions, $\xi_x$ and $\xi_y$, are given by
\[
\xi_x=\eta_R, \quad \xi_y=-\eta_I.
\]
We assume a quasi-stationary equilibrium, so that the characteristic time variation of equilibrium quantities ($t_{ch}$) is much longer than the period of kink oscillations ($P$) and introduce the small parameter $\epsilon$ so that $P= \epsilon t_{ch}$. Using the definition of the period of oscillations we can write
\begin{equation}
\frac{L(\mu_0\rho_{ch})^{1/2}}{B_0}=\epsilon t_{ch}\Longrightarrow B=\epsilon^{-1}\frac{L(\mu_0\rho_{ch})^{1/2}}{t_{ch}},
\label{eq:1.6}
\end{equation}
meaning that we can introduce a scaled magnetic field, so that ${\tilde B}_0=\epsilon B_0$. As a result, the equation describing the dynamics of the kink oscillations can be written as
\begin{equation}
\rho_i\left(\frac{\partial }{\partial t}+U_i\frac{\partial }{\partial z}\right)^2\eta+ \rho_e\left(\frac{\partial }{\partial t}+U_e\frac{\partial }{\partial z}\right)^2\eta-\frac{2\epsilon^{-2}{\tilde B}^2}{\mu_0}\frac{\partial^2\eta}{\partial z^2}=0.
\label{eq:1.7}
\end{equation}
The above equation must be solved subject to the standard boundary conditions, i.e.
\[
\eta(z=0, z=L)=0.
\]
Following the solution method proposed by Ruderman (2011a) we will solve Eq. (\ref{eq:1.7}) using the Wentzel-Kramers-Brillouin (WKB) method (see, e.g. Bender and Orsz\'ag 1987) and assume that the solution of the equation will be of the form
\begin{equation}
\eta=\sum_{k=0}^{\infty}\epsilon^kS_k(z,t)\exp\left[\frac{i}{\epsilon}\Phi(t)\right]
\label{eq:1.8}
\end{equation}
In the first order of approximation (often called geometric optics), Eq. (\ref{eq:1.7}) reduces to 
\begin{equation}
\frac{\partial^2 S_0}{\partial z^2}+\frac{\Omega^2}{{\tilde c_K}^2}S_0=0,
\label{eq:1.9}
\end{equation}
where 
\[
\Omega=\frac{d\Phi(t)}{dt}, \quad {\tilde c_K}^2=\frac{2{\tilde B}^2}{\mu_0(\rho_i(z,t)+\rho_e(z,t))}.
\]
Equation (\ref{eq:1.9}) must be solved subject to the boundary condition $S_0=0$ when $z=0$ and $z=L$. Equation (\ref{eq:1.9}) together with the line-tying condition form an eigenvalue problem in which $\Omega$ is the eigenvalue and $\Omega^2$ is a real function. 

In the next order of approximation (also called the approximation of physical optics), Eq. (\ref{eq:1.7}) reduces to
\begin{equation}
\frac{\partial^2 S_1}{\partial z^2}+\frac{\Omega^2}{{\tilde c_K}^2}S_1=\frac{2i\Omega}{{\tilde c_K}^2}\left[\frac{\partial S_0}{\partial t}+\frac{S_0}{2\Omega}\frac{d\Omega}{dt}+\frac{\rho_iU_i+\rho_eU_e}{\rho_i+\rho_e}\frac{\partial S_0}{\partial z}\right],
\label{eq:1.10}
\end{equation}
that has to be solved subject to the boundary condition $S_1(z=0,z=L)=0$. The boundary-value problem determining $S_1$ has a solution only when the RHS of Eq. (\ref{eq:1.10}) satisfies the compatibility condition, i.e. the orthogonality to $S_0$. After multiplying the RHS of Eq. (\ref{eq:1.10}) by $S_0$ and integrating with respect to $z$ in the interval $(0,L)$ we obtain (similar to Ruderman 2011a) that the compatibility condition reduces to
\begin{equation}
\omega\int_0^L\frac{S_0^2}{c_K^2}dz=const.,
\label{eq:1.12}
\end{equation}
where
\[
\omega=\epsilon^{-1}\Omega, \quad {c_K}=\epsilon^{-1}{\tilde c_K}.
\]
As a consequence, the dynamics of kink oscillations in coronal loops is fully described by the system of equation (\ref{eq:1.9}) and (\ref{eq:1.12}). In deriving Eq. (\ref{eq:1.12}) we took into account the mass conservation equation, relating the plasma flow and its density
\[
\frac{\partial\rho}{\partial t}+\frac{\partial (\rho U)}{\partial z}=0.
\] 

Let us now discuss a special case when the density depends on time only. This case would correspond to an initial expansion of the loop when the height of the loop is less than the scale-height (assuming expansion into an isothermal 1 MK corona, this height would correspond to 47 Mm), In this case, we may expect that the amplitude of oscillations increases as the loop expands. Indeed, it is easy to show that the amplitude of oscillations behaves like
\begin{equation}
A(t)=A(0)\left(\frac{c_K(t)}{c_K(0)}\right)^{1/2}=A(0)\left(\frac{L(t)}{L(0)}\right)^{1/2}
\label{eq:1.14}
\end{equation}
where $A(0)$, $c_K(0)$ and $L_0$ are the amplitude of oscillations, the kink speed and the length of the loop at $t=0$. 

A particular case to discuss is when the expansion of the loop occurs linearly with time and we write that 
\begin{equation}
L(t)=L_0+v_rt
\label{eq:1.15}
\end{equation}
where $L_0$ is the length of the loop at the initial time, i.e. at the TR level ($L_0=3\pi$ Mm) and $v_r$ is the rising speed, here assumed constant. Accordingly, the kink speed becomes
\begin{equation}
c_K^2=c_{Kf}^2\exp\left[\frac{L_0+v_rt}{\pi H}\sin\frac{\pi z}{L_0+v_rt}\right]
\label{eq:1.16}
\end{equation}
where 
\[
c_{Kf}^2=\frac{2B^2}{\mu_0\rho_f(1+D)}.
\]

Returning to the general case, the equations describing the dynamics of kink oscillations are 
\begin{equation}
\frac{\partial^2S_0}{\partial z^2}+\frac{\omega^2}{{c_K}^2}S_0=0, \quad \omega\int_0^L\frac{S_0^2}{c_K^2}dz=const.
\label{eq:1.17}
\end{equation}
Let us introduce a new set of dimensionless quantities
\begin{equation}
\xi=\frac{z}{L_0}, \;\;\tau=\frac{v_rt}{L_0}, \;\;{\tilde \omega}=\frac{\omega L_0}{c_{Kf}}, \;\;\tilde {S}_0=\frac{S_0}{L_0}.
\label{eq:1.18}
\end{equation}
In the new variables the equations to be solved transform into
\begin{equation}
\frac{\partial^2\tilde{S}_0}{\partial \xi^2}+\exp\left[\frac{-L_0(1+\tau)}{\pi H}\sin\frac{\pi \xi}{1+\tau}\right] {\tilde \omega}^2{\tilde S}_0=0, 
\label{eq:1.19}
\end{equation}
and
\begin{equation}
{\tilde \omega}\int_0^{1+\tau}{\tilde S}_0^2 \exp\left[\frac{-L_0(1+\tau)}{\pi H}\sin\frac{\pi \xi}{1+\tau}\right] d\xi=const.
\label{eq:1.20}
\end{equation}
that should be solved subject to the boundary conditions $S_0(\xi=0; \xi=1+\tau)=0$. The solution of the system (\ref{eq:1.19}) - (\ref{eq:1.20}) can be found numerically using, e.g. the shooting method. In Fig. 2 we display first the variation of periods of oscillations of the fundamental mode and its first harmonic for three different values of $H$ in terms of the dimensionless time variable, $\tau$. 
\begin{figure}
\centering
\includegraphics[width=\columnwidth]{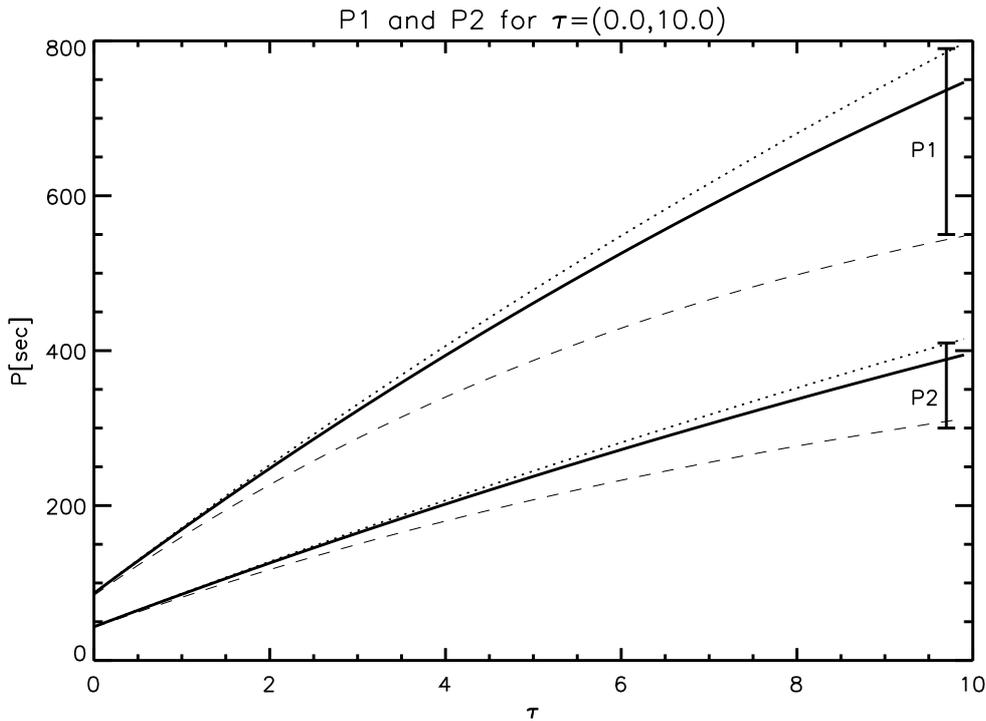}
\caption{The variation of the periods of the fundamental mode and its first harmonic with the dimensionless time variable $\tau$ for three different values of stratification: H=70.5 Mm (dotted line), H=47 Mm (solid line), and H=23.5 Mm (dashed line). }
\label{fig2}
\end{figure}
The bands for each period are clearly labelled in the figure. The three distinct value of periods were obtained for three values of scale-height keeping the initial length of the loop at 3$\pi$ Mm. The dotted line corresponds to a scale-height of 70.5 Mm, which, assuming a plasma in hydrostatic equilibrium, would correspond to a plasma temperature of 1.5 MK. The solid line is plotted for a loop expanding into a corona where the constant scale-height is 47 Mm, that would correspond to a 1MK hot plasma. Finally, the dashed line stands for an expansion of the loop into a plasma where the density scale-height is 23.5 Mm, corresponding to a plasma temperature of 0.5 MK. The two bands for the periods clearly show that the two oscillations are differently affected by the expansion, i.e. change in the length of the loop. This is also obvious in Fig. 3 where we plotted the ratio of the periods of the fundamental and first harmonic as a function of the time variable, $\tau$.  
\begin{figure}
\centering
\includegraphics[width=\columnwidth]{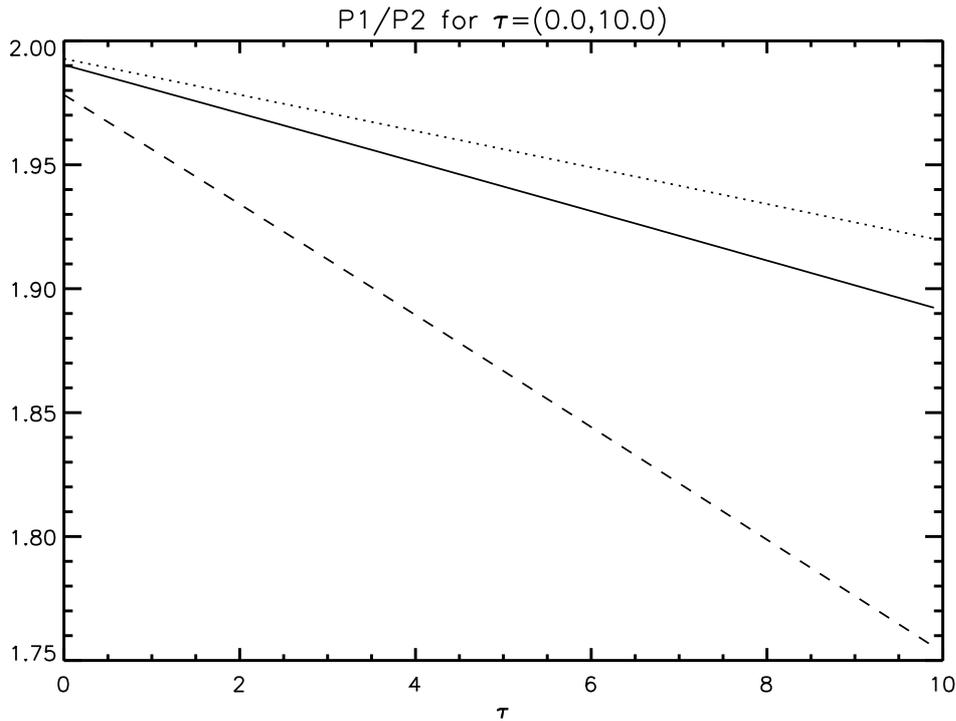}
\caption{The variation of the $P_1/P_2$ period ratio with respect to the dimensionless parameter $\tau$, for a loop expanding in the solar corona with persisting semi-circular shape. The meaning of each line-style is identical to Fig. 2.}
\label{fig3}
\end{figure}
The plot clearly shows that the oscillations of a loop expanding into a "hot" plasma (i.e. large scale-height) are the least affected, but in all three cases the period ratio decreases with time. We should note here that the periods shown in Fig 2 do not start at the value of 2 because at the start of their expansion through the solar corona, they are already stratified and the least stratified is the case that corresponds to $H=1.5$ MK. Let us make a final note: in observations, the identification of periods in coronal loops is a dynamical process. i.e. in EUV the intensity is measured in one location (or mega-pixel to reduce errors) for a long time-period. The duration of observation vary, and mostly they are driven either by the availability of the instrument or by the limited life-time of oscillations before they are damped by, e.g. resonant absorption, although observations show that not all loop oscillations damp (probably these oscillations are maintained by a constant lateral buffeting). Typically, for  damped oscillations, the detection time is a few periods with a range of 6.7-90 minutes (Aschwanden 2004). In terms of the dimensionless quantity $\tau$, this range would correspond to $\tau=0.64-8.6$. Although the lower limit is too small to count in the effect of expansion, a duration of $\tau=8.6$ would add an important effect in studying the transverse kink oscillations. Fig. 3 shows that for one given value of internal structuring, the value of the period ratio can change also because of the expansion of the loop. Although the periods shown in Fig. 2 display a monotonic increase with the parameter $\tau$, in reality these values will saturate, the saturation occurring faster for those modes that propagate in a highly structured plasma (e.g. for the case of H=23.5 Mm, the saturation value of periods is about 400 seconds and the saturation starts at about $\tau=20$). In addition, the period ratio for all cases discussed here tend to same value (near 1) for large values of $\tau$. {We need to mention here the very important fact that the effect of expansion on the periods of oscillations and the period ratio is relevant only in the expansion phase of the loop.}

{Analytical solutions of the Eqs. (\ref{eq:1.19}) can be obtained for the limiting case of a loop at the beginning of its expansion through the solar corona, i.e. small values of $\tau$ (see Appendix). The results confirm the tendency of periods to increase with time and of the period ratio to decrease with time. }

\section{Non-circular emergence}

In reality, the expansion of a loop in the empty corona so that the semi-circular shape is preserved is unlikely since the foot points have to move in a much denser plasma than the apex of the loop. That is why in this section we will assume that the expansion rate in the vertical direction is larger than the expansion of footpoints in the horizontal direction. The expansion still starts at the TR level where the shape of the loop is semi-circular and assume that the process remains isothermal. As a result of different expansion rates, the loop evolves so that the shape becomes more elliptical. The properties of transverse loop oscillations in an elliptical coronal loop was studied recently by Morton and Erd\'elyi (2009) assuming that the semi-elliptical shape is reached in the emerging stage, before reaching a semi-circular shape. Although their topic is related to the research of the present study, the problem of expansion is a dynamical process that should be treated accordingly. They found that the difference in $P_1/P_2$ period ratio between the circular and elliptical shape is up to 6\%.  

Since the dynamics is going to be different over the two directions, it is more convenient to introduce a polar coordinate system in which 
\begin{equation}
x=a(t)\cos \theta, \quad z=b(t)\sin \theta,
\label{eq:2.1}
\end{equation}
with $ \dot{a}(t)<\dot{b}(t)$, the overdot denotes the derivative with respect to time and the length of the loop is covered by the parameter $\theta$ that varies now between 0 and $\pi$. It is more convenient to use the coordinates along the loop, $s$, therefore the dynamics of transverse kink oscillations is described by 
\begin{equation}
\rho_i\left(\frac{\partial }{\partial t}+U_i\frac{\partial }{\partial s}\right)^2\eta+ \rho_e\left(\frac{\partial }{\partial t}+U_e\frac{\partial }{\partial s}\right)^2\eta-\frac{2B^2}{\mu_0}\frac{\partial^2\eta}{\partial s^2}=0,
\label{eq:2.2}
\end{equation}
A key parameter in our discussion is going to be the arc-length that is defined as
\begin{equation}
\frac{\partial s}{\partial \theta}=\sqrt{a(t)^2\sin^2\theta+b(t)^2\cos^2\theta}=\alpha(\theta,t).
\label{eq:2.3}
\end{equation}
 In order to solve Eq. (\ref{eq:2.2}) we would need to express the density as function of $s$ and $t$. However, it turns out that it is much easier to deal with the variable $\theta$ instead. Therefore we express the derivatives in the governing equation as
 \begin{equation}
 \frac{\partial}{\partial s}=\frac{1}{\alpha}\frac{\partial}{\partial \theta}, \;\; \frac{\partial^2}{\partial s^2}=\frac{1}{\alpha^2}\frac{\partial^2}{\partial \theta^2}-\frac{\sin2\theta(a(t)^2-b(t)^2)}{2\alpha^4}\frac{\partial}{\partial\theta}.
 \label{eq:2.4}
 \end{equation}
Assuming again a quasi-stationary equilibrium ,similar as in the previous section, and introducing the small parameter, $\epsilon$, the governing equation for transverse kink oscillations can be written as
\[
\frac{\partial^2\eta}{\partial t^2}+\frac{2}{\alpha}\left(\frac{\rho_iU_i+\rho_eU_e}{\rho_i+\rho_e}\right)\frac{\partial^2\eta}{\partial \theta \partial t}+\frac{1}{\alpha^2}\left[\frac{\rho_iU_i^2+\rho_eU_e^2}{\rho_i+\rho_e}-\right.
\]
\[
\left.\frac{2\epsilon^{-2}B_0^2}{\mu_0(\rho_i+\rho_e)}\right]\frac{\partial^2\eta}{\partial \theta^2}-\left[2\frac{\rho_iU_i+\rho_eU_e}{\rho_i+\rho_e}\frac{a\dot{a}\sin^2\theta+b\dot{b}\cos^2\theta}{\alpha^3}+\right.
\]
\[
\left.+\frac{\rho_iU_i^2+\rho_eU_e^2}{\rho_i+\rho_e}\frac{\sin2\theta(a(t)^2-b(t)^2)}{2\alpha^4}-\right.
\]
\begin{equation}
\left. 2\frac{\epsilon^{-2}B_0^2}{\mu_0(\rho_i+\rho_e)}\frac{\sin2\theta(a(t)^2-b(t)^2)}{2\alpha^4}\right]\frac{\partial \eta}{\partial \theta}=0,
\label{eq:2.5}
\end{equation}
that has to be solved subject to the boundary conditions $\eta(\theta=0,\theta=\pi)=0$. 

Again, we will solve this equation using the WKB approximation presented earlier and suppose a solution identical to Eq. (\ref{eq:1.8}). In the first order of approximation we obtain the equation
\begin{equation}
\frac{\partial^2 S_0}{\partial \theta^2}-\frac{\sin2\theta(a^2-b^2)}{2\alpha^2}\frac{\partial S_0}{\partial \theta}+\frac{\alpha^2\Omega^2}{{\tilde c}_K^2}S_0=0,
\label{eq:2.6}
\end{equation}
together with the usual boundary conditions at the two ends of the loop. In the next order of approximation we obtain
\[
\frac{\partial^2 S_1}{\partial \theta^2}-\frac{\sin2\theta(a^2-b^2)}{2\alpha^2}\frac{\partial S_1}{\partial \theta}+\frac{\alpha^2\Omega^2}{{\tilde c}_K^2}S_1=
\]
\begin{equation}
\frac{2i\alpha\Omega}{{\tilde c}_K^2}\left[\frac{\partial S_0}{\partial t}+\frac{\alpha S_0}{2\Omega}\frac{\partial \Omega}{\partial t}+\frac{\rho_iU_i+\rho_eU_e}{\rho_i+\rho_e}\frac{\partial S_0}{\partial \theta}\right].
\label{eq:2.7}
\end{equation}
This equation has to be solved subject to the boundary condition $S_1(\theta=0,\theta=\pi)=0$. Again, the equation for $S_1$ will have solution if the right-hand side of the above equation satisfies the compatibility condition, i.e. the orthogonality to $S_0$. Following the same solving procedure as presented earlier, it is easy to show that the condition reduces to
\begin{equation}
{\omega}\alpha\int_0^{\pi}\frac{S_0^2}{c_K^2}\;d\theta=0
\label{eq:2.8}
\end{equation}
Therefore, the system of equations (\ref{eq:2.6}) and (\ref{eq:2.8}) will determine completely the dynamics of the expanding coronal loop. 

Let us assume that at $t=0$ the loop is semi-circular and its length is $L_0$. In order to reproduce the different movement over the two directions, we introduce two different expansion speeds in the horizontal ($v_h$) and vertical ($v_v$) direction, so that $v_h<v_v$. Again, we suppose that the motion occurs linearly in time and write the dynamics over the two axes as
\begin{equation}
x=\left(\frac{L_0}{\pi}+v_ht\right)\cos\theta, \quad z=\left(\frac{L_0}{\pi}+v_vt\right)\sin\theta.
\label{eq:2.9}
\end{equation}
Let us introduce a new set of dimensionless quantities
\begin{equation}
\tau=\frac{v_vt}{L_0}, \quad {\tilde \omega}=\frac{\omega L_0}{c_{Kf}},\quad {\tilde \alpha}=\frac{\alpha}{ L_0}, \quad U=\frac{v_h}{v_v}, \quad  {\tilde S_0}=\frac{S_0}{L_0}.
\label{eq:2.10}
\end{equation}
In the new variables, the governing equations become
\[
\frac{\partial^2{\tilde S}_0}{\partial \theta^2}+{{\tilde \alpha}^2{\tilde \omega}^2}\exp\left[-\frac{L_0(1+\tau\pi)}{\pi H}\sin\theta\right]{\tilde S}_0+
\]
\[
\frac{\sin\theta\cos\theta[\tau^2\pi^2(1-U^2)+2\tau\pi(1-U)]}{1+\tau^2\pi^2(\sin^2\theta+U^2\cos^2\theta)+2\tau\pi(\sin^2\theta+U\cos^2\theta)}\times \]
\begin{equation}
\frac{\partial{\tilde S}_0}{\partial \theta}=0,
\label{eq:2.11}
\end{equation}
and
\begin{equation}
{\tilde \omega}{\tilde \alpha}\int_0^{\pi}(\frac{{\tilde S}_0^2}{{\bar c_K}^2}\;d\theta=0,
\label{eq:2.12}
\end{equation}
where ${\bar c_K}=c_K/v_v$.
\begin{figure}
\centering
\includegraphics[width=\columnwidth]{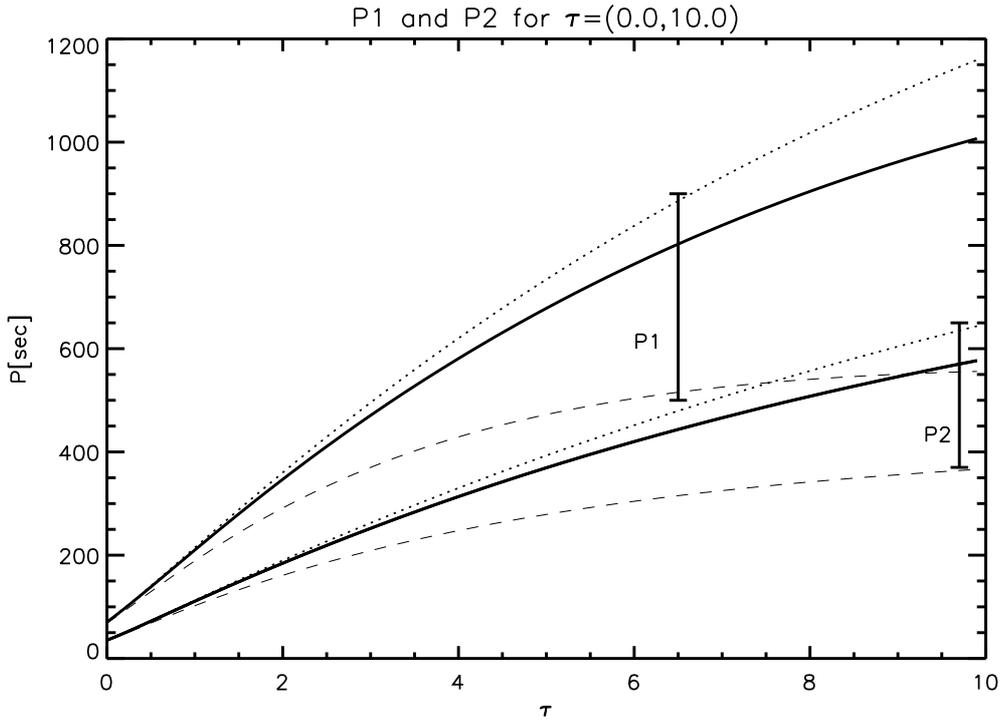}
\caption{The same as in Fig. 2, but here we assume that the expansion of the loop occurs such that the loop evolves into a loop with a semi-elliptical shape. The meaning of different line-styles is identical to Fig 2.}
\label{fig5}
\label{fig4}
\end{figure}
Figure 4 displays the evolution of the period of oscillations for the fundamental mode and its first harmonic for three different values of $L_0/H$, similar values as used in the previous section. Comparing the findings in Figs 2 and 4, the effect of the expansion into an elliptical shape compared to the constant semi-circular shape is evident. As time progresses the period of oscillations tend to a higher value for elliptical shape, however this conclusion is more true for the fundamental mode. The period of the fundamental mode corresponding to an expansion in a solar corona where scale height is only 23.5 Mm (corresponding to a temperature of 0.5 MK in a loop in hydrostatic equilibrium) saturates rather quickly. 
\begin{figure}
\centering
\includegraphics[width=\columnwidth]{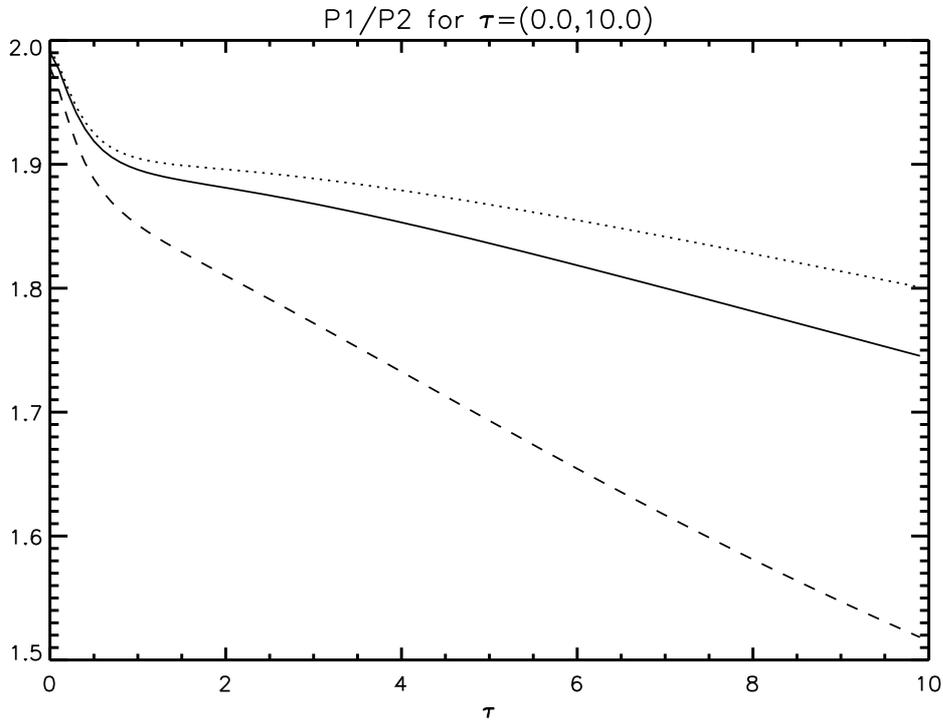}
\caption{The same as in Fig. 3, but here we assume that the expansion of the loop occurs such that the loop evolves into a loop with a semi-elliptical shape. The meaning of different line-styles is identical to Fig 2.}
\label{fig5}
\end{figure}

A more significant change is evident when comparing the $P_1/P_2$ period ratio of the expansion into an elliptical shape shown in Fig. 5 to the variation of the period ratio for a semi-circular shape. For the same time interval the decrease of the period ratio is much more significant, and similar to the previous case, the period ratio is more affected for the case of strong stratification, i.e. small H.

\section{Conclusions}

The solar corona is a very dynamical environment where changes in the dynamical state of the plasma and field occur on all sort of time scales. In the present study we combined for the first time two kinds of dynamical events: the time-evolution of a coronal loop through its expansion into the "empty" corona and the transverse kink oscillations of coronal loops. The emergence and expansion of a coronal loop is a very complex phenomenon, but here we reduced our model to a simplified process, where the expansion is solely described by the change in the length of the loop with an associated temporal equilibrium density variation.

The governing equation for kink oscillations was solved in the WKB approximation when the boundary conditions are time-dependent. As expected, due to the change in the length of the loop, the amplitude and periods of oscillations increase with time, however, the period ratio of the fundamental mode and its first overtone decreases. This last physical parameter is of paramount importance for the remote determination of density structuring of coronal loops with the help of seismological approaches. In the first instance we regarded the loop to have an initial semi-circular shape that is maintained through the expansion phase. Later, this restriction was lifted based on the natural assumption that the expansion into the vertical direction (i.e. in the direction of density decrease) occurs much easier than in the horizontal direction. In this limit, the loop evolves into a semi-ellipse, with the major axis in the vertical direction. Comparing the results of the two approaches it is clear that the behaviour of the period ratio is rather sensitive to the geometrical shape of the loop, a more significant drop in the $P_1/P_2$ ratio being achieved in the second case. Although our numerical result were obtained for three different structuring degrees (measured by the ratio of the initial loop length to the density scale-height) it is also evident that both the temporal change in the loop length and the stratification will have the same effect upon the period ratio resulting in a mutual amplification of the effect.

Our model predicts that the amplitude of oscillations increases with time, however due to the particular choice of density, damping processes were neglected. Once the density is allowed to vary also in radial direction, according to the theory of resonant absorption (e.g. Goossens et al. 1992, Rudeman and Roberts 2002), loops will damp very quickly with the resonant position displaying a steady motion due to the change of the length of the loop. The amplitude of oscillations can also be damped due to the cooling of the plasma (Morton et al. 2010), an effect that was also neglected here. In an expanding loop, the growth of the amplitude due to emergence and decay of amplitude due to resonant damping or cooling will be competing processes and the competition between these two effects will be discussed in a forthcoming paper.

\acknowledgements

IB acknowledges the financial support by NFS Hungary (OTKA, K83133).
The authors are grateful to Prof M.S. Rudemman and Dr. V. Archontis for their help and comments.


\appendix
\section{Solutions to the wave equation in the case of the loop at the beginning of its expansion }

An interesting insight into the character of the solution of the governing equation can be obtained analytically in the limiting case of 
\[
\zeta=L_0(1+\tau)/\pi H\ll1, 
\]
i.e. we restrict ourself to the first part of the emergence into the solar corona. In this case the argument of the exponential function in Eq. (\ref{eq:1.19}) can be expanded (keeping only the first two terms) and the equation to be solved reduces to
\begin{equation}
\frac{\partial^2\tilde{S}_0}{\partial \xi^2}+ {\tilde \omega}^2{\tilde S}_0=-\sin\frac{\pi \xi}{1+\tau}\zeta{\tilde S}_0, 
\label{eq:1.21}
\end{equation}
Since we are looking for periodic solutions and we expect that both the amplitude and frequency will depend on time, we will employ the Poincar\'e-Lindstedt method (Meirovitch 1970) to find corrections to the eigenfunctions and eigenfrequencies (it can be easily shown that this method is similar to the re-normalization technique used by Ballai et al 2007 in the case of dispersive shocks). We are looking for solutions in the form of series and write
\begin{equation}
 {\tilde S}_0=\sum_{k=1}^{\infty} \zeta^k{\tilde S}_0^{(k)}, \quad {\tilde \omega}=\sum_{k=1}^{\infty} \zeta^k{\tilde \omega}^{(k)},
 \label{eq:1.22}
 \end{equation}
where the functions $S_i$ are periodic functions. We first concentrate on the fundamental mode. After inserting the expansions (\ref{eq:1.22}) into the governing equation (\ref{eq:1.21}), we collect terms proportional to subsequent powers of $\zeta$. In the first order of approximation, collecting terms $\sim {\cal O}(\zeta^0)$, results in
 \begin{equation}
 \frac{\partial^2{\tilde S}_0^{(0)}}{\partial \xi^2}+{\tilde \omega^{(0)2}}{\tilde S}_0^{(0)}=0.
 \label{eq:1.23}
 \end{equation} 
 Solving this equation subject to the mentioned boundary conditions, yields in the case of the fundamental mode 
 \begin{equation}
 {\tilde S}_0^{(0)}\sim\sin \frac{\pi\xi}{1+\tau}, \quad {\tilde \omega}^{(0)}=\frac{\pi}{1+\tau}.
 \label{eq:1.24}
 \end{equation}
 In the next order of approximation (i.e. terms $\sim {\it O}(\zeta)$) we obtain 
  \[
 \frac{\partial^2{\tilde S}_0^{(1)}}{\partial \xi^2}+{\tilde \omega^{(0)2}}{\tilde S}_0^{(1)}=-\frac{\pi^2}{(1+\tau)^2}\sin^2\frac{\pi\xi}{1+\tau}-
 \]
 \begin{equation}
- \frac{2\pi}{1+\tau}{\tilde \omega}^{(1)}\sin\frac{\pi\xi}{1+\tau},
 \label{eq:1.25}
 \end{equation} 
 that has to be solved subject to the boundary conditions
 \begin{equation}
 {\tilde S}_0^{(1)}(\xi=0,\xi=1+\eta)=0.
 \label{eq:1.26}
 \end{equation}
The last term in Eq. (\ref{eq:1.25}) will cause secular growth of the solution, therefore rendering the solution $S_1$ non-periodic. To suppress this possibility we choose ${\tilde \omega}_1=0$. As a result, the solution of Eq. (\ref{eq:1.25}) together with the boundary conditions is simply written as
\begin{equation}
{\tilde S}_0^{(1)}\sim \sin\frac{\pi \xi}{1+\tau}+\frac32\cos\frac{\pi \xi}{1+\tau}-\frac16\left(3+\cos\frac{2\pi \xi}{1+\tau}\right)
\label{eq:1.26.1}
\end{equation}
In the next order of approximation we collect terms ${\cal O}(\zeta^2)$ and obtain
\[
\frac{\partial {\tilde S}_0^{(2)}}{\partial \xi}+{\tilde \omega^{(0)2}}{\tilde S}_0^{(2)}=-{\tilde \omega^{(0)2}}{\tilde S}_0^{(1)}\sin\left(\frac{\pi\xi}{1+\tau}\right)-
\]
\begin{equation}
2{\tilde \omega^{(0)}}{\tilde \omega^{(1)}}{\tilde S}_0^{(0)}.
\label{eq:1.26.2}
\end{equation}
Using the expression of ${\tilde \omega^{(0)}}$, ${\tilde S}_0^{(0)}$ and ${\tilde S}_0^{(1)}$ determined earlier, the RHS of the above equation can be written as
\[
RHS=\left[\frac{\pi^2}{3(1+\tau)^2}-\frac{2\pi{\tilde \omega}^{(2)}}{1+\tau}\right]\sin\frac{\pi\xi}{1+\tau}+A\sin^2\frac{\pi\xi}{1+\tau}+
\]
\begin{equation}
B\sin\frac{2\pi\xi}{1+\tau}+C\sin\frac{3\pi\xi}{1+\tau},
\label{eq:1.26.3}
\end{equation}
where the coefficients of higher harmonics ($A$, $B$ and $C$) are not needed for our discussion. In order to prevent non-periodic behaviour we need to impose the condition that the coefficient of the first term is zero, leading to 
\begin{equation}
{\tilde \omega}^{(2)}=\frac{\pi}{6(1+\tau)}.
\label{eq:1.26.4}
\end{equation}
As a result, the eigenfunction and eigenfrequency of fundamental mode oscillations can be written as
\[
{\tilde S}_0=\sin\frac{\pi\xi}{1+\tau}+\zeta\left[\frac32\cos\frac{\pi\xi}{1+\tau}-\frac16\left(3+\cos\frac{2\pi\xi}{1+\tau}\right)\right],
\]
\begin{equation}
\tilde{\omega}=\frac{\pi}{1+\tau}+\zeta^2\frac{\pi}{6(1+\tau)},
\label{eq:1.26.5}
\end{equation}
meaning that the change in the frequency due to expansion is a second order effect. It is easy to show that the period of the fundamental mode in this approximation can be written as
\begin{equation}
P_1=\frac{12(1+\tau)}{6+\zeta^2},
\label{eq:1.26.6}
\end{equation}
proving the increase of the period $P_1$ with $\tau$ seen in Fig 2.

Repeating the same method for the first harmonic, where the eigenfunction and eigenfrequency in the zeroth-order approximation are
\[
{\tilde S}_0=\sin\frac{2\pi\xi}{1+\tau}, \quad {\tilde \omega}^{(0)}=\frac{2\pi}{1+\tau},
\]
we obtain that
\begin{equation}
{\tilde \omega}=\frac{2\pi}{1+\tau}+\zeta^2\frac{2\pi}{15(1+\tau)},
\label{eq:1.26.7}
\end{equation}
meaning that the period of the first harmonic in this approximation behaves like
\begin{equation}
P_2=\frac{15(1+\tau)}{15+\zeta^2},
\label{eq:1.26.8}
\end{equation}
showing an increasing tendency with respect to $\tau$. Now, using Eqs. (\ref{eq:1.26.6}) and (\ref{eq:1.26.8}) we can calculate the period ratio of the fundamental mode and its first harmonic as
\begin{equation}
\frac{P_1}{P_2}\approx 2\left(1-\frac{\zeta^2}{60}\right)=2\left[1-\frac{L_0^2(1+\tau)^2}{60\pi^2H^2}\right],
\label{eq:1.26.9}
\end{equation}
so that the change in the period ratio is very small but decreases with $\zeta$.

\end{document}